# A model-driven approach for composing SAWSDL semantic Web services

**Fatima-Zahra Belouadha, Hajar Omrana and Ounsa Roudiès**

**Computer Science Department, Mohammadia School of Engineers (EMI), Mohammed V[th] University-Agdal
BP. 765 AV. Ibn Sina Agdal, Rabat, Morocco**

## Abstract

Composing Web services is a convenient means of dealing with complex requests. However, the number of Web services on the Internet is increasing. This explains the growing interest in composing Web services automatically. Nevertheless, the Web services' semantics is necessary for any dynamic composition process. In this article, we present an MDA approach to develop and compose SAWSDL semantic Web services. To model Web services, we use a UML profile which is independent of the description standards. The SAWSDL interface files are generated by using transformation rules. To model the behavior of a composite Web service and generate its executable BPEL file, we use the BPMN notation in a platform of modeling and implementing business process. The main contribution of this work is the easy and extensible solution to a model-driven development of the semantic atomic and composite Web services.

## 1. Introduction

The appearance of the component paradigm and the evolution of the Internet have fostered the emergence of Web services. This paradigm, born of the need for B2B exchange and communication, is essential as a promising technology for the interoperability of the heterogeneous systems in particular. Web services are a new emerging standard for integration and B2B exchanges. Despite the growing supply of Web services on the Internet, their composition to meet a complex request remains a crucial objective that helps improve the business productivity.

At their emergence, Web services were only described by their functional capabilities using the WSDL (Web Services Description Language) standard. This standard is known as the ancestor of all Web services description languages. It was used to describe most of the existing Web services. Today, other languages such as OWL-S (Web Ontology Language for Web Services), WSML (Web Service Modeling Language), WSDL-S (Web Services Semantics) and SAWSDL (Semantic Annotations for WSDL) add a semantic layer to the descriptions of

Web services. Developing semantic Web services has a crucial interest for particularly enabling their dynamic composition.

SAWSDL is a recent W3C recommendation [1]. Its main advantage remains in its extensibility and compatibility with the WSDL standard. We think that the development of the tools which help generate SAWSDL files is important. These tools provide the semantics of Web services which are required by dynamic composition processes. They also help map WSDL to SAWSDL and conversely. This is the main motivation of this work. Our objective is to propose a MDA (Model-Driven Architecture) solution for the implementation of the atomic and composite semantic Web services. The idea is to develop executable Web services by automatically generating SAWSDL and BPEL (Business Process Execution Language) files.

In section 2, we present a metamodel of semantic atomic and composite Web services. This metamodel is independent of the description standards and the ontology languages. The section 3 describes a UML profile for this metamodel. In sections 4 and 5, we focus on the semantic elements of the SAWSDL metamodel and the transformation rules we developed to realize the mapping to SAWSDL. The section 6 describes how we implement a composite Web service. It presents our approach to model the composite Web service behavior and generate the corresponding BPEP file. The section 7 constitutes an evaluation of this work by comparing it with some related works. Our conclusions and perspectives are given in section 8. To clarify our work, we transversally present a comprehensive example.

## 2. Independent PIM metamodel

In the MDA approach, the model is the core of any work. Our proposal is based on a metamodel of the composite semantic Web services. This metamodel is independent of the SAWDL language. It lies in an abstract level and conceives a Web service as a business service. As illustrated in figure 1, we model a Web service as a business service that realizes one interface. This interface is a set of operations. Each operation can have input and output parameters as well as faults. Each parameter belongs to a specific data type.

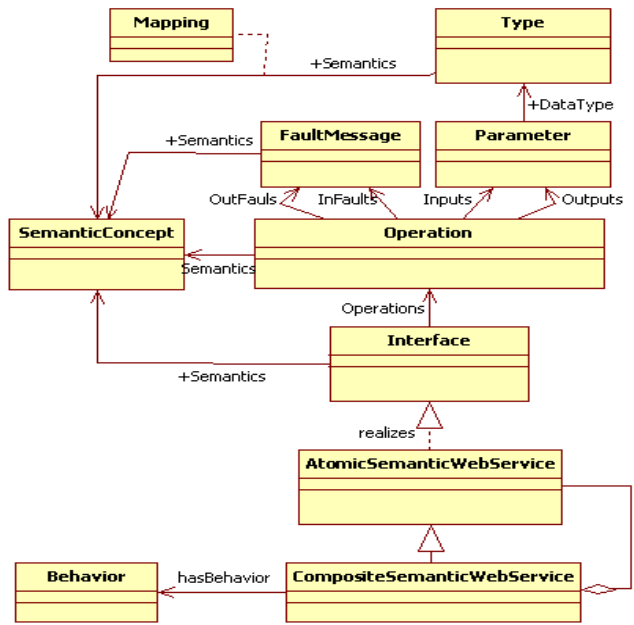

Fig. 1 SAWSDL Independent metamodel of composite semantic Web services.

The semantic layer of a Web service is specified using semantic concepts. The interface, the operation and the type of each parameter can be associated with a semantic concept. The semantic concept can be an ontological concept. For example, the interface can be linked to a taxonomy that provides information on the category of service (e.g., parapharmaceutical sale, travel service, etc.). The operation is attached to a semantic concept which provides information on its functionality (e.g., list of products, products prices, etc.). The parameter type must also be associated with a concept that provides information on the parameter's semantics. In particular, an operation parameter is associated with its data type. The mapping between two types must be specified. For example, if we consider a data type *CardType* which is represented by a string containing the number, the type and the expiration date of the card separated by dots and its equivalent ontological concept represented as a structure of three separated fields, it will be necessary to specify the mapping between this structure and the given string. In our metamodel, the two-way mapping between data types of a given parameter and its semantic concept are specified using two attributes *LoweringSchema* and

*LiftingSchema*. These attributes are properties of an association class named *Mapping* in figure 1.

A semantic composite service is an aggregation of two or more atomic or composite services. It inherits the properties of an atomic service. However, it has a behavior. A behavior provides information about the coordination and the orchestration of the composed operations. The metamodel presented in this section corresponds to the PIM (Platform Independent Platform) level of the MDA approach.

## 3. UML Profile

The developer needs to design his own model of Web services. For this purpose, we have created a UML profile corresponding to the metamodel proposed in section 2. This profile is supposed to be easy for the software developers. It should not require any knowledge of Web services standards and semantic models languages. Today, UML is the successor of the Merise method which is the most known language of software designers. It contains stereotypes and tagged values. Stereotypes are attached to model elements to convey the meaning of those elements. Tagged values are name/value pairs. They are attached to model elements in order to supply additional information which is needed in the transformation process.

Our profile contains stereotypes and tagged values. The stereotypes which we use are *AtomicSemanticWebService*, *CompositeSemanticWebService*, *Behavior*, *in param*, *Type*, *SemanticConcept*, *Mapping*, etc. They correspond to the elements described in section 2. These stereotypes extend the UML class as well as other UML elements. For example, the stereotype *AtomicSemanticWebService* extends, in addition, the UML interface which encapsulates a set of properties and methods (operations). The stereotype *SemanticConcept* extends the operation, the parameter and the interface of the UML standard. This makes it possible to apply this stereotype to these elements and thus to specify tagged values providing semantic information about them. Especially, to semantically annotate a Web service, our profile provides tagged values such as the URI of a semantic concept and the mapping schemas (*LoweringSchema* and *LiftingSchema*). These tagged values respectively annotate a parameter by the URI of its semantic concept and the URIs which provide information about its data types mapping. Tagged values can also be used to model the information about the Web service binding (e.g., endpoints, message exchange protocols, etc.). However, this type of information is not necessary at the modeling level. It is used during the execution of the service and can be specified at PSM (Platform Specific Model) level.

To illustrate the elements of the proposed profile, we consider the example of a composite Web service for electronic sale *ElectronicSale*. The service receives a credit card and a price as inputs and withdraws the sum corresponding to this price if the card is valid. It returns an error message if the card is invalid or if the regulation could not be done. This service is composed of two atomic Web services *CardValidate* and *ePayment*. The service *CardValidate* receives a card and checks if it is valid, otherwise it returns an error message. The service *ePayment* withdraws the cash and returns an error message if a problem occurs. A card is represented by a structure composed of the card number, the card type and the expiration date.

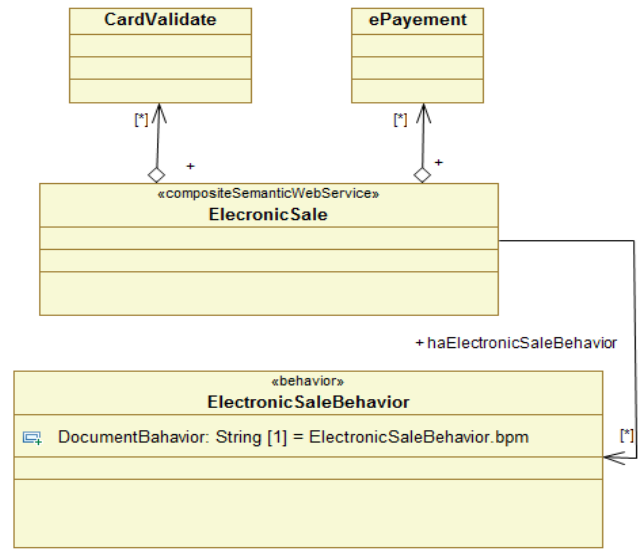

Fig. 3 UML interface model of the composite Web service *ElectronicSale*.

## 4. SAWSDL Metamodel

SAWSDL is an extension of the WSDL syntactical description. It defines an annotation mechanism that allows to semantically describe Web services in terms of concepts provided by a domain ontology, and it provides these semantic annotations embedded into WSDL documents [2].

Transforming languages-independent models (PIM models created using our UML profile) to generate SAWSDL files requires the specification of the PSM metamodel. This metamodel represents the target format to be transformed according to SAWSDL specifications. In this sense, we presented in a previous work a SAWSDL metamodel of Web services based on WSDL 1.1 [3]. For the sake of optimization and compliance with the new W3C recommendations, we refine and improve this metamodel to be compatible with the new WSDL version 2.0 and the new recommendations. In WSDL 2.0, the element *message* including the message parts is replaced by the operation inputs and outputs. The element *PortType* is replaced by the element *interface*. The binding (providing technical information such as the transport protocols) is no longer encapsulated in the binding tag. We focus, in this section, on how SAWSDL integrates semantic annotations into a WSDL file.

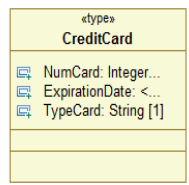

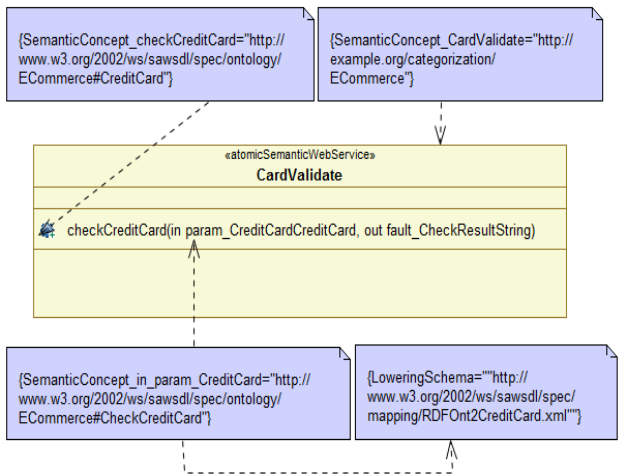

Fig. 2 UML interface model of the atomic semantic Web service *CardValidate*.

The figure 2 illustrates an extract of the interface model of the atomic semantic Web service *CardValidate* given as example. This model was created using the profile proposed in this section. The semantic annotations are described by tagged values that relate elements to semantic concepts (e.g., URI of a domain ontology).

The figure 3 illustrates an extract of the interface model of the composite semantic Web service *ElectronicSale*. The composite pattern is used to model the relationship between the service *ElectronicSale* and the couple constituted of the services *CardValidate* and *ePayment*.

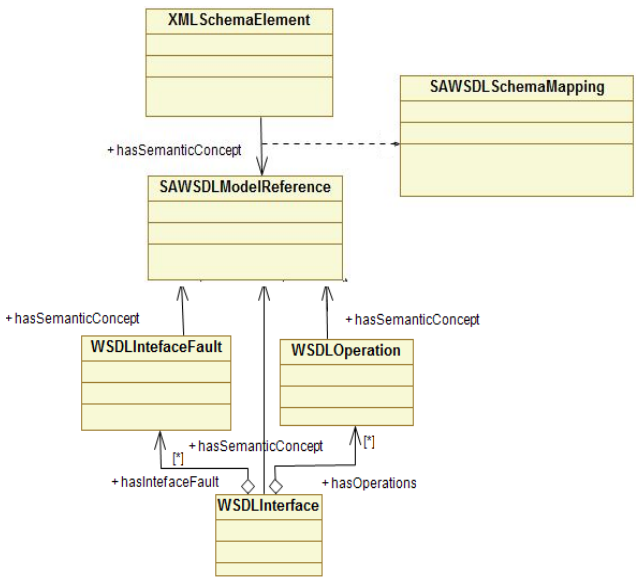

Fig. 4 Extract of the SAWSDL metamodel of semantic Web services.

As illustrated in figure 4, our SAWSDL metamodel includes a set of metaclasses which comply with SAWSDL and WSDL 2.0. According to WSDL 2.0, an interface file is described by a tag *WSDLDescription* which is an aggregation of interfaces. The interface is modeled as a metaclass named *WSDLInterface*. Each interface is realized by a Web service. It is an aggregation of operations and their faults which are represented by the metaclasses *WSDLOperation* and *WSDLInterfaceFault*. The metaclass *XMLSchemaElement* is the data representation of simple or complex types according to XML schemas. These types are associated with operation parameters. To semantically annotate the elements describing a Web service, the metaclasses *SAWSDLModelReference* and *SAWSDLSchemaMapping* are used. The metaclass *SAWSDLModelReference* associated with the metaclasses (the interface, the operation, the faults and the XML Schema) used to describe a Web service provides information about their semantic concepts. The metaclass *SAWSDLSchemaMapping* is an association class which provides information on the tow-way mapping between an XML schema and a parameter type.
Finally, the metamodel presented in this section correspond to PSM (Platform Specific Model) level of the MDA approach.

## 5. Mapping to SAWSDL model

Transformation rules are necessary to realize the mapping from the PIM models (e.g., our language independent model) into PSM ones (e.g., our SAWSDL model). For this purpose, we used the ATL (ATLAS Transformation Language) language to implement these rules [4]. This language is the ATLAS research group answer to the OMG MOF/QVT RFP. It is a model transformation language specified both as a metamodel and as a textual concrete syntax. ATL comply with the MDA approach and uses a repository to store and manipulate the source and target metamodels. It executes the transformation according to defined transformation rules.

The figure 5 shows an example of the transformation rules we developed to enable the mapping between the UML interface model and the SAWSDL model. It illustrates the code of the rule *SemanticInterface*. When executed, this rule creates instances of the SAWSDL metaclass *WSDLInterface* for each class whose stereotype is *AtomicSemanticWebService* in the interface model. The name of an instance is set according to the name of its corresponding class in the UML interface model.

```
rule SemanticInterface {

from a: UML!Class (a.stereotype ->

exists(e|e.name='AtomicSemanticWebService'))

to b: SAWSDL!Interface

( name <- a.name),
}
```

Fig. 5 Example of transformation rules.

The table 1 shows a list of mapping between the elements of the UML profile and the metaclasses of the SAWSDL metamodel. For example, the stereotype *AtomicSemanticWebservice* is mapped to the class *WSDLInterface*. We consider a Web service as a business service which implements an interface. The stereotype *SemanticConcept* is mapped to the class *ModelReference*. It provides information about the URI of an ontological concept. The tagged value *LiftingSchema* is mapped to *SAWSDLLiftingSchema* which is a specialization of *SAWSDLSchemaMapping*. The figure 6 shows an extract of the SAWSDL file corresponding to the atomic semantic Web service *ValidateCard*. The SAWSDL file was automatically generated by transformation.

Table 1: Examples of Mapping between the UML profile and the SAWSDL metamodel

| UML Profile element | Type | SAWSDL Metaclass |
|---|---|---|
| AtomicSemanticWebService | Stereotype | WSDLInterface |
| AtomicSemanticWebService's Method | Stereotype | WSDLOperation |
| in param | Stereotype | WSDLInput |

| out param | Stereotype | WSDLOutput |
| in fault | Stereotype | WSDLInfault |
| out fault | Stereotype | WSDLOutfault |
| Type | Stereotype | XMLSchemaElement |
| SemanticConcept | Stereotype | SAWSDLModelReference |
| Mapping | Stereotype | SAWSDLSchemaMapping |
| LoweringSchema | Tag Value | SAWSDLLoweringSchema |
| LiftingSchema | Tag Value | SAWSDLLiftingSchema |

```
<xs:element name="CreditCard"
    sawsdl:modelReference="http://www.w3.org/2002/ws
        /sawsdl/spec/ontology/ECommerce#CheckCreditCard"
    sawsdl:loweringSchemaMapping="http://www.w3.org/2002/
        ws/sawsdl/spec/mapping/RDFOnt2CreditCard.xml">
    <xs:complexType>
     <xs:sequence>
      <xs:element name="NumCard" type="xs:integer" />
      <xs:element name="ExpirationDate" type="xs:integer" />
      <xs:element name="TypeCard" type="xs:string" />
     </xs:sequence>
    </xs:complexType>
   </xs:element>
</wsdl:types>

<wsdl:interface name="CardValidate"
    sawsdl:modelReference="http://example.org/
        categorization/ECommerce">
    <wsdl:fault name="CheckResult" />

    <wsdl:operation name="checkCreditCard"
        pattern="http://www.w3.org/ns/wsdl/in"
    sawsdl:modelReference="http://www.w3.org/2002/ws/sawsdl/
        spec/ontology/ECommerce#CreditCard">
    <wsdl:input element="CreditCard" />
    <wsdl:outfault ref="CheckResult" messagelabel="CheckResult"/>

    </wsdl:operation>
 </wsdl:interface>
</wsdl:description>
```

Fig. 6 Extract of the SAWSDL file of *ValidateCard* service.

# 6. Modeling the composite Web service behavior

In our SAWSDL-independent metamodel, a composite Web service has a behavior. This behavior provides information about the orchestration and the coordination of the set of operations which compose the Web service. We think that the modeling of a composite Web service behavior seems to be the same task as modeling a business process. Both adopt the same way of organizing the process execution.

In this perspective, we opted for the BPMN notation (Business Process Modeling Notation). This new standard is more convenient for the orchestration of business processes than the UML activity diagrams. BPMN is poised to become the new standard for modeling business processes and Web services [5]. Business processes, designed using the BPMN standard, can be directly mapped to any business modeling executable language for immediate execution.

Both of the standards BPEL4WS (Business Process Execution Language for Web Services) and BPML (Business Process Modeling Language) provide specifications for dataflow, messages, events, business rules, exceptions and transactions. The orchestration of Web services could be defined as a workflow of activities constituting a business process. Control constructs, such as the sequence, the parallel split, the synchronization, the exclusive choice and the multiple choices as well as the messages and the events must be considered. In particular, the execution of composite Web services could require a sequential execution of certain components. Other components could, for example, be executed in parallel depending of the validity of a given condition. In the case of a composite Web service for travel, this service could book a ticket for a given date. If the reservation is confirmed, a parallel execution of the services hotel and car reservation could follow.

The figure 7 shows how we model the behavior of the composite service *ElectronicSale* using BPMN. This work was performed using Intalio designer tool. The behavior of the service *ElectronicSale* is modeled by the sequential orchestration of the services *CardValidate* and *ePayment*. This sequence depends on the validity of a given card.

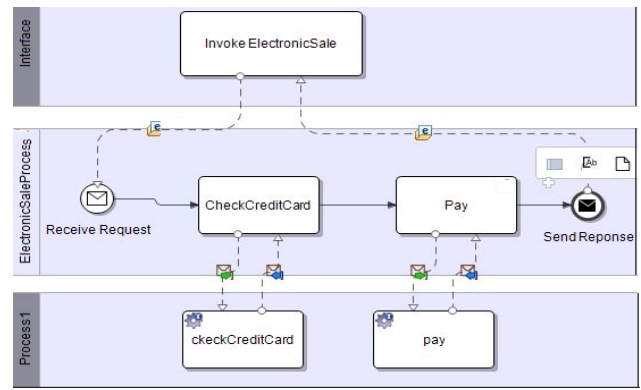

Fig. 7 BPMN model of the composite Web service *ElectronicSale*.

When the behavior modeling's step is achieved, it is possible to generate the SAWSDL and the BPEL files. Extracts of the SADWSL and BPEL files contents related to the composite Web service *ElectronicSale* are shown in figures 8 and 9. The BPEL file enables the BPEL server to execute the composite Web service.

```
<wsdl:import namespace="http://example.com/ElectronicSale/
     ElectronicSaleProcess"
     location="ElectronicSaleProcess-ElectronicSaleProcess.wsdl"/>
<wsdl:import namespace="http://us.intalio.com/CardValidate/"
     location="Service/CardValidate.wsdl"/>
<wsdl:import namespace="http://us.intalio.com/ePayment/"
     location="Service/ePayment.wsdl"/>
<pnlk:partnerLinkType
     name="ElectronicSaleProcessAndProcess1ForPortTypeCardValidateSoap
     <pnlk:role name="Process1_for_ElectronicSaleProcess"
     portType="tns:CardValidateServiceSoap"/>
</pnlk:partnerLinkType>
<pnlk:partnerLinkType
     name="ElectronicSaleProcessAndProcess1ForPortTypeePaymentSoapPlk
     <pnlk:role name="Process1_for_ElectronicSaleProcess"
     portType="tns:ePaymentServiceSoap"/>
</pnlk:partnerLinkType>
<pnlk:partnerLinkType name="ElectronicSaleProcessAndInterface">
     <pnlk:role name="ElectronicSaleProcess_for_Interface"
     portType="ElectronicSaleProcess:ForInterface"/>
</pnlk:partnerLinkType>
```

Fig. 8 Extract of SAWSDL file content related to the composite Web
service *ElectronicSale*.

```
<bpel:variables>
<bpel:variable name="thisReceive_RequestRequestMsg"
     messageType="this:Receive_RequestRequest"/>
<bpel:variable name="thisReceive_RequestResponseMsg"
     messageType="this:Receive_RequestResponse"/>
<bpel:variable name="tnsCheckCreditCardRequestMsg"
     messageType="tns:checkCreditCardpIn"/>
<bpel:variable name="CardValidateServiceCheckCreditCardResponseMsg"
     messageType="Process1:checkCreditCardResponse"/>
<bpel:variable name="tnsPayRequestMsg"
     messageType="tns:paySoapIn"/>
<bpel:variable name="tnsPayRequestMsg"
     messageType="tns:paySoapIn"/>
<bpel:variable name="ePaymentPayResponseMsg"
     messageType="Process1:payResponse"/>
<bpel:variable name="ePaymentPayResponseMsg"
     messageType="Process1:payResponse"/>
</bpel:variables>

<bpel:sequence>
<bpel:receive
     partnerLink="electronicSaleProcessAndInterfacePlkVar"
     portType="this:ForInterface" operation="Receive_Request"
     variable="thisReceive_RequestRequestMsg" createInstance="yes"
bpmn:label="Receive Request" name="Receive_Request"
bpmn:id="_iS9B8NIXEd6NDptfMmxqPg"></bpel:receive>
```

Fig. 9 Extract of BPEL file content related to the composite Web service
*ElectronicSale*.

# 7. Evaluation and discussion

Our work is founded on a MDA approach to generate
SAWSDL files and statically compose Web services using

models. A lot of interesting work has been performed in
this area. Formal models were adopted to describe
particularly the functional aspect of Web services.
Predicate path expressions, process algebra, finite states
automate, Petri nets and semantic graphs, all were
explored for this purpose [6, 7, 8, 9, 10]. UML models
were also adopted in several works. They prove to be
simple and better known by the most developers. Provost
proposes a WSDL 1.0 dependent UML metamodel to
describe the atomic Web services [11]. This metamodel
integrates WXS schema (W3C XML Schema) and object
oriented types. Gronmo and Skogan present a WSDL
independent UML model to generate the OWL-S
composite Web services [12]. They propose an extension
of the UML activity diagrams according to control
patterns in order to specify Web services behaviors [13].
Gronmo and al. use a UML profiles package [14],
including the OWL ontology profile defined by Duric [15],
to describe the OWL-S composite Web services semantics.
Timm and Gannod use an OWL-S dependent UML profile
to model atomic processes [18]. They use extended UML
activity diagrams to specify composite processes [19].
Control constructs (e.g., decision nodes, fork nodes, etc.)
are specified using The OCL (Object Constraint
Language) language before being transformed.

Using the UML activity diagrams to describe the behavior
of composite services does not seem the appropriate
decision. In particular, the composite Web service
execution follows the guidelines of its BPEL file that
reflects its behavior. However, UML does not define any
execution metamodel for business processes modeled with
it [16]. Moreover, in spite of its expressivity, OWL-S is
not aligned with the existing Web services standards. It
also assumes the use of OWL to represent ontologies [17].

Other works proposed semantic models of Web services
using ontology based approaches. As a first attempt in this
area, Narayanan and McIlraith provide a model-theoretic
semantics as well as a distributed operational semantics
that can be used for simulation, validation, verification,
automated composition and enactment of Web services
[20]. They describe the capabilities of Web services using
the DAML-S DAML+OIL ontology. A first-order logical
language is used to define the semantics for a subset of
DAML-S. The service descriptions are encoded in the
Petri Net formalism. In this work, the authors provide
decision procedures for Web service simulation,
verification and composition. Rao, Küngas and al.
describe an approach for the automatic composition of
DAML-S semantic Web services [21]. They use Linear
Logic (LL) theorem proving. The services are presented
by axioms and proofs. A process calculus is used to
formally present a composite service. A semantic reasoner is exploited to relax the matching process. Both works [20, 21] have as objective to ensure a dynamic composition of Web services using ontology, logical languages and formal models.

Rajasekaran and al. propose METEOR-S tools for source code annotation and semantic Web service generation [22]. The user can use a semantic Web service designer GUI to manually associate Web services parameters with ontological concepts in the source code level. The annotations are represented in Java, by using the meta-tag feature of the new j2sdk, jdk1.5. The annotated source code is used by a semantic description generator module to generate annotated WSDL 1.1, WSDL-S and OWL-S files. A source code implements the functional aspect of a given service and not its non-functional properties. We think that its use for Web services annotation allows the annotation of some services parameters existing in the source code, such as inputs, outputs and constraints. However, this manner does not make it possible to annotate non-functional properties such as the quality of service or the security level.

SAWSDL is a recent recommendation of W3C. Verma and Sheth present primary tools for semantic annotation of Web services using SAWSDL [23]. In particular, SAWSDL4J is an implementation of the SAWSDL specification. Radiant is an Eclipse plug-in for creating SAWSDL and WSDL-S service interfaces. Developers can use the SAWSDL4J API to write source codes in order to create SAWSDL services. They can also add annotations to existing WSDL documents via OWL ontology using Radiant's GUI. Easy annotation can be done by Drag and Drop. However, users must master the proposed API and have knowledge of the SAWSDL language to create new Web services. Moreover, the services composition is not a purpose of the proposed tools.

Gomadam and al. propose a declarative approach to dynamically compose SAWSDL Web services [24]. The requester specifications concerning a required process can be different from the process specifications created by the provider. To solve the problem of the processes heterogeneities, the authors use a planner and a data mediator. Data and functional requirements are specified using semantic templates. The mediator converts queries messages into the format of the input message of a relevant operation which meets the partner requirements. A GraphPlan algorithm based on the task specification is used to generate a BPEL file. The proposed approach solves a non-trivial real-world problem regarding the 2006 SWS Challenge mediation scenario version 1 [25]. It allows to automatically compose SAWSDL Web services.

However, it currently supports only sequence, AND-split and loop workflow patterns.

## 8. Conclusions and perspectives

The EDI standards and the technologies of information exchange in a distributed environment (CORBA, DCOM and RMI) are not sufficient to ensure at the same time the systems interoperability and the simplicity of implementation. Thanks to their standards based on XML, Web services enable today to meet better these needs. Composing Web services brings an added-value and allows meeting unexpected requests. Consequently, the composition of Web services is one of the most active research fields. In this paper, we present a UML model-based approach, which is aligned with the new standards, to describe and compose semantic Web services. Our profile allows to create new Web services, or to extend the existing ones (after converting the description files to UML models) with semantic properties. The semantic annotation is embedded into generated SAWSDL files using transformations rules. Description elements can be semantically described in terms of concepts provided by semantic models. No conditions on the language used to construct these semantic models (e.g., ontologies) are required. Users can manually compose semantic services using the BPMN standard through Intalio designer GUI.

A lot of work has been performed in the semantic Web services area. However, our approach is different in terms of methodology and advantages. First, we use a simple metamodel for Web services which is independent of the WSDL standard and the semantic models languages (such as OWL). This helps software developers overcome the complexity of the standards used for Web services implementation. They do not need to learn any language. They also can annotate services parameters with semantic concepts regardless of ontology languages. Besides, our description language-independent metamodel is extensible. It can be extended to consider elements, such as preconditions, postconditions and effects, utilized in other languages. This will allow generating and mapping different Web services files such as OWL-S and WSML documents. Second, we utilize the BPMN standard to model the composite Web service behavior instead of the UML activity diagrams. The BPMN model of the given service is directly mapped to an executable BPEL file. No further steps are required for modeling the fully executable processes. Third, we use UML profiles as a means of describing functional aspects of Web services. This makes it possible to extend the services description to non-functional properties (such as the security level or the quality of service) by simply adding new profiles to our package. Besides, our approach is aligned with the last recommendations of the W3C. The proposed profiles and transformations allow generating Web services whose description files comply with the extensible SAWSDL standard. Thus, it will be naturally and quite possible to introduce, in the description files, non-functional properties using the WS-Policy standard.

Our work lies in the category of the static composition. However, the generated Web services can be used by automatic processes of discovery, selection and dynamic composition of Web services. Especially, concepts from the semantic models, which are referenced from within WSDL elements as annotations, can be represented using formal models or languages. This helps disambiguate the description of Web services during matchmaking processes.

The development of an example of composite services using the MDA approach enabled us to implement and validate our model. However, to evaluate the quality of the service composition, we envision to apply our approach to concrete examples (e.g., existing semantic Web services) or to a problem of SWS challenge [25]. This allows us to compare the results provided by atomic services with those provided by the corresponding composite service. The initial objective of this work is to generate semantic Web services. However, to use these semantics in automatic processes, it will be necessary to use matchmaking algorithms such as the algorithms proposed in [26, 27]. The matchmaking processes help discover similar concepts of a domain ontology which annotate Web services.

As perspectives of this work, we are actually working on the extension of our metamodel in order to describe non-functional properties using WS-Policy standard. We are considering integrating other languages, like OWL-S and WSML, used to describe semantic Web services. Our metamodel will be a combination of different components used by these languages. A reflection on this work was already highlighted in our previous works [3, 28, 29]. We are also working on the reverse transformations rules which will convert a SAWSDL file to a UML model. We are considering extending the transformations to other languages. These operations will enable mapping between various languages and the easy integration of a semantic layer into existing Web services like the WSDL services.

At the end of this work, we have been able to conclude that the use of a Web services description languages-independent model is sufficient to generate their interface files. Nevertheless, a model, which complies with a standard description language, is primitive to develop mapping rules from the independent model. Such a model would also be used for the opposite conversion (of the interface file of an existing Web service into the language-independent interface model). Our second conclusion concerns the modeling of the composite Web services' behavior. The BPMN notation meets better this need than the UML activity diagrams. It enables to automatically generate the executable BPEL files.

## Acknowledgments

The Authors of this article thank Mr. Mounire Benhima, Director General of PowerAct Consulting, Manager of Systems Quality Management Department and Consultant for his technical advice in process management.

**F. Z. Belouadha** Doctorate degree in Computer Science with distinction in 1999; Former Assistant chief of the Computer Science Department at the Mohammadia School of Engineers (EMI); Assistant Professor at the Computer Science Department-EMI; Best paper award at SIIE'08 Conference; Recognition award at IEEE International AICCSA'09 Conference; 9 recent publications papers between 2007 and 2009; Ongoing research interests: Semantic composite Web services, Pervasive information systems/m-services, Business intelligence, MDA .

**H. Omrana** Engineer degree in Computer Science in 2003; DESA degree in Computing Networks, Telecommunications and Multimedia in 2006 with distinction (Major of promotion); PhD Student in Computer Science; Chief of the Computer Science department in a service company; Ongoing research interests: Semantic Web services composition, Quality of Web services.

**O. Roudiès** Doctorate degree in Computer Science in 1989, PhD in Computer Science in 2001; Former chief of the Computer Science Department at the Mohammadia School of Engineers (EMI); Chief of Computer Science field at EMI; Professor at the Computer Science Department-EMI; Co-Editor of the e-ti Journal; 15 recent publications papers between 2008 and 2009; Ongoing research interests: SI, composition, Web services, patterns, Quality.